\documentclass[reprint,superscriptaddress]{revtex4-1}
\pdfoutput=1
\usepackage{amsmath, amssymb}
\usepackage{graphicx}
\usepackage{color}
\usepackage{hyperref}
\usepackage{float}

\graphicspath{{figures/}}

\begin{abstract}
Multi-component self-assembly mixtures offer the possibility of encoding multiple target structures with the same set of interacting components. Selective retrieval of one of the stored structures has been attempted by preparing an initial state that favours the assembly of the required target, through seeding, concentration patterning or specific choices of interaction strengths. This may not be possible in an experiment where on-the-fly reconfiguration of the building blocks to switch functionality may be required. In this paper, we explore principles of inverse design of a multi-component self-assembly mixture capable of encoding two competing structures that can be selected through simple temperature protocols. We design the target structures to realise the generic situation in which one of targets has the lower nucleation barrier while the other is globally more stable. We observe that to avoid the formation of spurious or chimeric aggregates, the number of neighbouring component pairs that occur in both structures should be minimal. Our design also requires the inclusion of components that are part only of one of the target structures, but we observe that to maximize the selectivity of retrieval, the component library itself should be maximally shared by the two targets. We demonstrate that temperature protocols can be designed which lead to the formation of either one of the target structures with high selectivity. We discuss the important role played by secondary aggregation products, which we term \emph{vestigial aggregates}. 
\end{abstract}

\begin{document}

\title{Temperature Protocols to Guide Selective Self-Assembly of Competing Structures}

\author{Arunkumar Bupathy}
\email{bupathy@jncasr.ac.in}
\affiliation{Theoretical Sciences Unit, Jawaharlal Nehru Centre for Advanced Scientific Research, Bangalore, India.}

\author{Daan Frenkel}
\email{df246@cam.ac.uk}
\affiliation{Department of Chemistry, University of Cambridge, Cambridge, UK.}

\author{Srikanth Sastry}
\email{sastry@jncasr.ac.in}
\affiliation{Theoretical Sciences Unit, Jawaharlal Nehru Centre for Advanced Scientific Research, Bangalore, India.}

\maketitle

\section*{Introduction}

Self-assembly is a fundamental manufacturing mechanism of Nature. 
Many meso-scale cellular structures required for biological functionality such as membranes, microtubules, actin fibrils, ribosomes, etc., are formed through self-assembly, often driven by non-equilibrium forces~\cite{Whitesides2002,Whitesides2002a}. 
Even though the cytoplasm contains thousands of molecular components, the various cellular structures self-assemble with remarkable precision, and may even share components~\cite{Kuhner2009}. However, in rare cases they can misassemble leading to impaired function or even diseases~\cite{Lansbury1999,Chiti2006}. Mechanisms for controlling synthesis in the cell, such as molecular chaperones and compartmentalisation of enzymatic action 
~\cite{chaperones,Jaiman2020,Yadav2020}, are widely studied. 

When designing self-assembling systems, one must contend with  unwanted ``off-target'' structures. 
It is therefore important to understand what causes misassembly, and how it can be avoided,  both at the design stage and during assembly. On the other hand, an ability to reuse the same building blocks to assemble different structures can be extremely useful to create smart materials that can change their functionality in response to an external stimulus. 
Often, such materials are designed to change their shape and functionality through conformational or morphological changes of their building blocks~\cite{Nguyen2011, Gong2014, Yoo2010, Whitelam2009a, Batista2010, Lewandowski2015, Sacanna2013}. 
However, it is also possible to have multiple structures that differ in the spatial arrangement of their building blocks. 
Multi-component mixtures not only allow for such a design~\cite{Murugan2015, Zhong2017, Bisker2018, Barish2009}, but can also form finite structures with arbitrary complexity~\cite{Halverson2013, Zeravcic2014}. 
Experiments with DNA bricks have shown a way of self-assembling complex structures with hundreds of distinct components~\cite{Ke2012, Ong2017}. Such an \textit{addressable assembly}, where each component and its position in the target structure are uniquely defined, is made possible because of the complementarity of the DNA hybridization process.

Here, we investigate the generic problem of designing two competing target structures of distinct shape, a feature which we show involves new non-trivial challenges. By designing the location of components and the strength of their interactions, we show that the nucleation behaviour of the target structures can be tuned such that either of them can be assembled using distinct time varying temperature protocols. Further, such a design shows that to avoid chimeric aggregation, the neighborhood of the individual components in the two structures should be maximally different. We show that the design and selective self assembly of the competing structure is aided by the inclusion of components that form part of only one of the two structures, but maximizing the selectivity requires that the component library be maximally shared by the targets.  

We begin by noting that even with a single target structure, multi-component systems assemble quite differently from classical nucleation due to the fact that the components need to bind in certain specific ways to form the correct structure~\cite{Cademartiri2015, Frenkel2015}.
Numerical evidence and theoretical analyses have shown that multi-component self-assembly proceeds via a non-classical nucleation process~\cite{Reinhardt2014, Jacobs2015a, Jacobs2016, Jacobs2015}. For successful assembly, the theories also predict a protocol that allows for slow nucleation followed by completion of growth at a lower temperature~\cite{Jacobs2015,Jacobs2015b}.
Due to incidental interactions between components there may also be numerous undesired ways in which they can aggregate, which increases the probability of formation of undesired structures. 
Thus, the designed interactions should be made sufficiently strong so as to offset this entropic effect~\cite{Hedges2014, Hormoz2011}.
Some  studies suggested that a narrow distribution of designed interactions is required for error-free self-assembly~\cite{Hedges2014, Hormoz2011}. 
However, other studies showed that variable bond strengths may improve the kinetics, and  diminish the competition between fragments that are incorporated at an early stage~\cite{Jacobs2015,Madge2018}.
To avoid the formation of off-target structures, it is important that the self-assembling structure can anneal during growth.
This implies that the assembly should take place under thermodynamic conditions where the growth is almost reversible~\cite{Jacobs2015, Evans2017}. As a consequence,  the range of thermodynamic parameters within which self-assembly can be made error-free is significantly reduced~\cite{Ong2017, Madge2015, Reinhardt2014, Sajfutdinow2018}.

Focusing now on strategies to design multiple target structures from the same building blocks, the two most pertinent questions are: (i) how to design the targets while avoiding misaggregation, and (ii) how to guide their self-assembly into specific  target structures. The simplest examples of distinct targets that form from the same building blocks are objects that have the same shape, but differ in the spatial arrangement of building blocks~\cite{Murugan2015, Zhong2017, Bisker2018}. 
Components that are neighbors in any of the multiple targets are assigned attractive interactions. 
If the interactions are of equal strength, then the retrieval of any specific target from the mixture requires a target-specific seeding procedure or concentration pattern.
Selective retrieval by strengthening a few bonds specific to the desired target has also been attempted~\cite{Murugan2015, Bisker2018}, in which case the system always favors the formation of one structure over others.

\section*{Self-assembly targets}
\begin{figure*}
\centering
\includegraphics[width=0.95\linewidth]{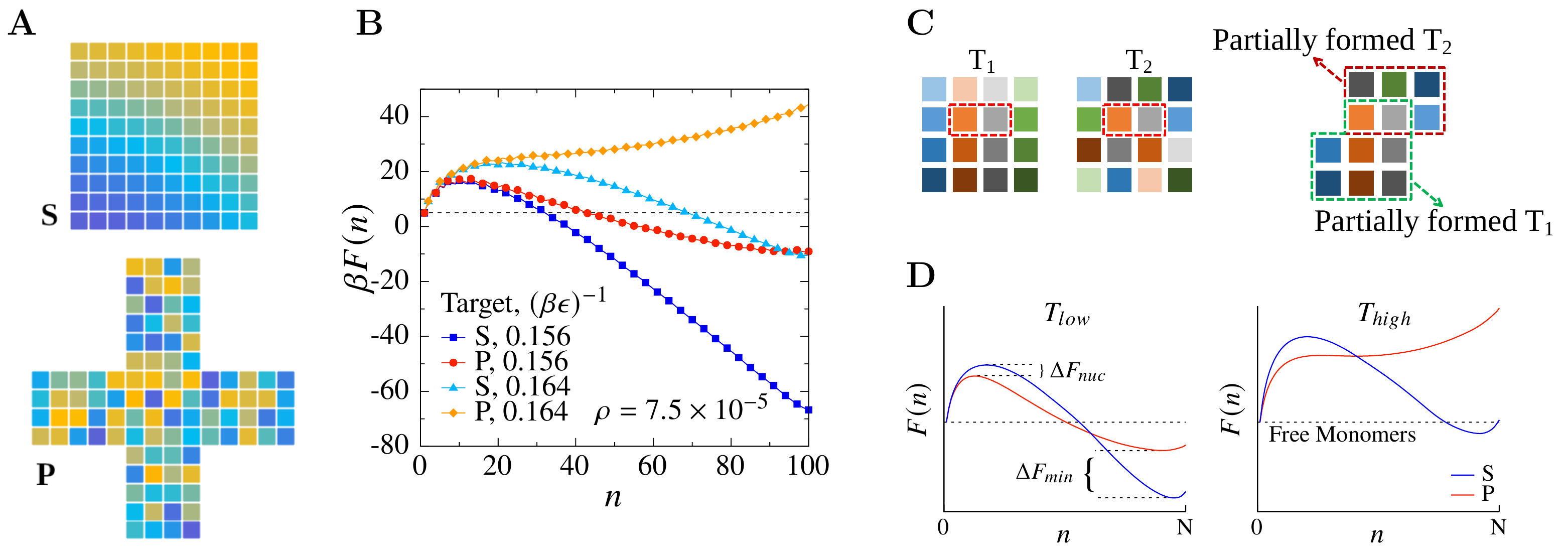}
\caption{Panel (A) shows the two structures that we use as design targets in this study. Each target is made of $100$ distinct components -- square blocks with four interacting faces and an orientation (not depicted for the sake of clarity). We shall change the composition of the targets later in our design. Panel (B) shows the free energy landscapes of the two targets at $k_BT/\epsilon=0.156$ and $0.164$ when all designed bonds are of equal strength $\epsilon$. (C) Illustration showing two example structures T$_1$ and T$_2$ that differ in their internal arrangement of components, except for one common bond as highlighted by the broken red rectangle. On the right, a possible chimeric aggregate that is part T$_1$ and part T$_2$ is shown. (D) Schematic showing a scenario where one target has the lower barrier while the other has the lower free energy minimum. As the temperature is increased, the structure with the lower barrier will become unstable first.}
\label{fig:design}
\end{figure*}

To explore the strategy for designing different structures from the same building blocks, we consider two structures (square (S) and plus (P)), defined on a two-dimensional square lattice (see Fig.~\ref{fig:design}(A)). 
Each target is composed of $N = 100$  square blocks, with four distinct interacting edges. 
Each component is represented with a distinct color, but the four possible orientations are not shown for the sake of clarity. 
To begin with, we consider S and P to be made of the same set of $100$ distinct components. The labelling of the blocks is arbitrary, and is done in sequential order in the S structure. These components are placed randomly in the P structure with random orientations.

Given the two targets, we need to specify interactions between the edges of the components such that they are stable. Representing by $k, l \in \{1,2,3...,4M\}$ a pair of component edges, where $M$ is the size of the component library, the interaction matrix $I$ that encodes both targets has the form
\begin{equation}
I_{kl}
\begin{cases}
 < 0, & \text{if the edges $k$ and $l$ are in}\\
 & \text{contact in S or P} \\
 = 0, & \text{otherwise.}
\end{cases}
\label{eq:combined_matrix}
\end{equation}

Initially, we assign equal strengths $-\epsilon$ to all the bonds 
and compute their free energy curves as a function of the aggregate size, using the method of Jacobs \textit{et al.}~\cite{Jacobs2015a, Jacobs2015}, which is outlined in the {\it SI Appendix}, Sec.~1.
In Fig.~\ref{fig:design}(B), we show the free energy curves for two temperatures, with all components having the same concentration $\rho = 7.5 \times 10^{-5}$ in the mixture, which corresponds to having $3$ copies of each component in a $200\times200$ lattice. While this is a relatively small system, it is sufficient to demonstrate the important features of our design. 


The differences in the connectivity of the two structures naturally gives rise to differences in their free energy landscapes. From Fig.~\ref{fig:design}(B), it is clear that we can devise a temperature protocol to selectively retrieve target S, since it can nucleate at the higher temperature shown where P is unstable, and the growth can be completed at the lower temperature ~\cite{Jacobs2015a, Jacobs2015}. 
However, we cannot define a protocol for the formation of P without the simultaneous formation of S. 
There could also be other spurious or chimeric aggregates, as explained below, and 
in general, it is not possible to predict such aggregates \textit{a priori}. 
However, we show below that we can avoid spurious aggregation by an appropriate choice of interactions and composition of the target structures. 
In the following sections, we discuss how to choose the target compositions, and our scheme for tuning the interactions that allows selective retrieval of either structure through different temperature protocols.

\section*{Choosing the target compositions}
\label{sec:composition}


Although it may be possible to design the structures with fewer components than N, we do not do so as larger libraries are better for kinetic accessibility of the structures~\cite{Hormoz2011, Trubiano2021}. Because the components are shared, the neighboring components with attractive interactions of a given component can be different in the two targets. 
Such promiscuous interactions could lead to assembly errors, because an exposed edge of a growing aggregate has multiple candidates for attachment. 
A possible strategy to minimize the number of aggregation paths due to promiscuous interactions is to maximize the number of components that share the share the same neighborhood in the two structures. 
However, such as strategy, while reducing the number of aggregation paths, causes chimeric structures to be more stable. 
We illustrate this in Fig.~\ref{fig:design}(C) which shows two structures T$_1$ and T$_2$, each made of the same set of $16$ distinct components, which differ in their internal arrangement of components. There is one pair of nearest neighbors that are shared by both the structures indicated by dashed red rectangles. 
On the righthand side, a possible chimeric aggregate is shown. 
The (stable) chimera is made of parts of T$_1$ and T$_2$ that are held together by two bonds. This shared motif acts as a stronger glue between the incompatible pieces than individual shared sites may be. 
This point is further detailed in {\it SI Appendix}, Sec.~2 and Sec.~3.  


We therefore minimize the number of shared bonds between the two structures.  
Additionally, the boundaries of the two targets need to be inert to prevent any  aggregation beyond completion of the desired structures. 
This is partly achieved by using the same set of components for the boundaries of each structure. 
However, since their boundaries can be of unequal lengths, there will be some blocks that have non-zero interactions. To avoid this, we choose a slightly larger component library to begin with, so that the extra components serve as additional boundary blocks. 
Before we proceed with the implementation, we discuss how to tune the interactions for protocol guided retrieval.


\section*{Tuning the interactions for targeted retrieval}
\label{sec:design}


Considering that a temperature range is available in which the target structures can form through a nucleation process, we generically have the possibilities that the free energy barrier and the free energy minimum for one structure is higher than for the other, or that one of the structures has a lower free energy barrier while having a higher free energy minimum when fully formed. In the former case, no obvious protocol exists to selectively form the structure with higher free energy barrier and minimum, whereas for the latter, such a protocol can be devised, as we discuss below. We thus consider the latter scenario. Further, as we describe, such a scenario can be realised through different approaches. 

The free energy curves for the situation we consider are shown schematically  in Fig.~\ref{fig:design}(D). At low temperatures, both the structures are stable. The structure P has the lower nucleation barrier but the structure S is globally more stable (or vice versa). As the temperature is increased, the structure P becomes unstable, while the structure S remains stable or metastable. At $T_{low}$, where P has a sizeable but surmountable nucleation barrier, we would nucleate P with a higher probability than S. If the nucleation barrier of S is sufficiently high, then we would rarely nucleate S so that we can grow P with a high degree of selectivity. On the other hand, at even lower temperatures, both S and P will be able to grow. Once they nucleate, we can then make P dissociate back into the mixture by increasing the temperature until P becomes unstable, thus selectively retrieving S. This gives us the temperature protocols for retrieving the targets.

We can achieve such a scenario by tuning the strengths of the individual bonds between the components (or the chemical potentials of the components, which we do not pursue here. Nevertheless, both approaches are illustrated in the SI Appendix Sec.~4. This is possible as long as there are enough bonds in either structure. Experimentally, the tuning of bond strengths can be achieved by varying the size or the number of attractive patches in the case of patchy colloids or by varying the strand lengths in the case of DNA bricks. 
Specifically, we require the product $\Delta F_{nuc} \times \Delta F_{min}$ to be negative and their magnitudes such that: (i) there is a sizeable difference in their nucleation rates and (ii) S remains stable for a sufficiently higher window of temperatures than P. In the following section, we perform the optimization of the target composition and the interaction strengths for our example structures.


\section*{Optimized target compositions and free energies}

In our example, target S has a boundary of length $36$ blocks and P has a boundary of length $50$ blocks. Using the boundary components from S to form the boundary of P, we would need an additional $14$ components to make the boundary of P fully inert, thus increasing the size of our component library $M$ to $114$. Consequently, only $86$ components are shared by both S and P, and each target has $14$ components that are unique to it.

We start with an initial assignment such that S is made of components $1$ to $100$, and P is made of components $15$ to $114$. We then iterate on the internal permutation of components and their orientations so as to minimize the cost function
\begin{equation}
C = n_{shared-bonds} + n_{active-boundary} + f\times \!\!\!\!\sum_{t \in \{S,P\}}\!\sum_{\substack{(\alpha,\beta)\\ shared}} r_{\alpha\beta}^t
\label{eq:compositioncost}
\end{equation}
where $n_{shared-bonds}$ is the number of shared bonds between the two targets, $n_{active-boundary}$ is the number of boundary components that have non-zero interactions, and $r_{\alpha\beta}^t = |\mathbf{r}_{\alpha}^t - \mathbf{r}_{\beta}^{t}|$ is the separation between a pair of components $\alpha$ and $\beta$ in target $t$, where $\alpha$ and $\beta$ are shared by both targets. $f \ll 1$ is a weight factor to ensure that the first two terms are minimized with higher priority. The third term ensures that the shared components form compact cores in either target. The reason for this shall be clear later, when we introduce the idea of vestigial aggregation as a chemical buffer.

\begin{figure}
    \includegraphics[width=0.95\linewidth]{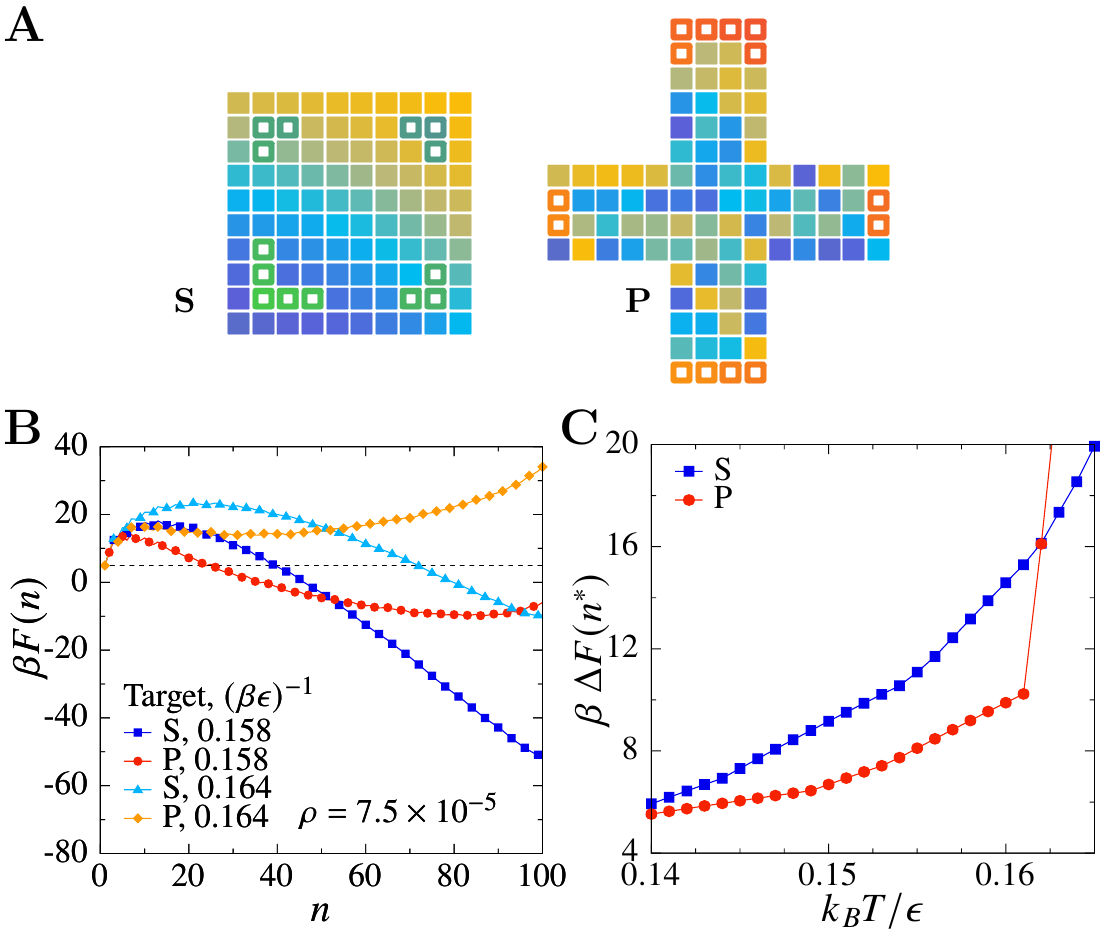}
    \caption{Panel (A) shows the target structures with optimized internal composition. The colors of the blocks represent the component type. The open squares of green and orange shades are unique to either targets while the other components are shared. Panel (B) shows their free energy curves after the bond strengths have been tuned as described in the text. Panel (C) shows the nucleation barriers of the two structures versus temperature, after the tuning.}
    \label{fig:optcomposition}
\end{figure}

The composition of the targets after such an optimization is shown in Fig.~\ref{fig:optcomposition}(A). The open green/white and orange/white blocks  are unique to either target while the other components are shared. 
The choice of a larger library ensures that there are no interacting boundary edges, and in our specific case there are no shared bonds between the two targets either, \textit{i.e.}, the first two terms of Eq.~(\ref{eq:compositioncost}) are zero in the optimized composition. 
Note that the composition shown here is not unique and that there may be many such equivalent arrangements.

Once the compositions are fixed, we tune the bond strengths. This is done by initializing the strength of all designed bonds to $-\epsilon$, and performing random updates on the individual bonds so as to minimize $\Delta F_{nuc} \times \Delta F_{min}$. The value of $\Delta F_{nuc}$ and $\Delta F_{min}$ are computed at a fixed temperature $k_B T / \epsilon = 0.156$ such that with all bonds at equal strength, both structures have a nucleation barrier of about $10k_B T$. For the free energy calculations, we took the free monomer concentration $\rho \equiv \rho_{\alpha} = 7.5 \times 10^{-5}$ for all component species $\alpha$, which is the value we use in our MC simulations. We tune the bond strengths until $|\Delta F_{nuc}| \approx 3 k_B T$, which gave $|\Delta F_{min}| \approx 42 k_B T$. 
While this method works well, it is not optimal for targeting the individual magnitudes of $\Delta F_{nuc}$ and $\Delta F_{min}$, and one might consider a different cost function.

In Fig.~\ref{fig:optcomposition}(B), we show the free energy curves after the optimization of the bond strengths. With the tuned interactions, the rate at which the nucleation barriers increase with temperature is different for the two structures, as shown in Fig.~\ref{fig:optcomposition}(C). 
Hence their nucleation times also grow apart with temperature, and this feature is crucial for our design. 
At $k_BT/\epsilon \approx 0.163$, P becomes unstable.


\section*{Retrieving the targets through temperature protocol}


For target P, the simulation temperature is chosen such that the estimated nucleation time of S is greater than P by a factor of at least $10$. The estimation of the nucleation times is described in SI Appendix Sec.~5. For target S, we start at a lower temperature where the nucleation time of S is no more than $4$ times that of P. We then ramp up to a temperature where P becomes unstable, which can be slightly higher than predicted by the free energy curves, and wait until the nuclei of P are dissolved. This is repeated multiple times to improve the yield of S, similar in spirit to kinetic proofreading mechanisms~\cite{Hopfield1974,Ninio1975}. 


We perform simulations with a fixed number of particles, instead of with a constant chemical potential. For selective retrieval of S, either protocol will work equally well. 
However, under conditions where P can be selectively nucleated, the assembled structure is not complete (see Fig.~\ref{fig:optcomposition}(B)), and lowering the temperature to complete the assembly, under constant chemical potential conditions, leads to the nucleation of S as well. 
While there may be other ways of addressing this issue, we consider closed systems -- which may  be easier to realise experimentally -- and devise a suitable protocol for such a set up.

Since we simulate the system in the canonical ensemble and allow for multiple copies of each target, the depletion of free monomers with time shifts the free-energy curves up. We compensate for this by lowering the simulation temperature. The temperature shift can be approximated by requiring that the free energy (divided by the temperature) of the targeted structure remains invariant with depletion ({\it SI Appendix}, Sec.~6), and is given by
\begin{equation}
\beta(t) = \beta(0) + \frac{N}{E(N)}\ln\frac{\rho(0)}{\rho(t)},
\label{eq:rescaledtemp}
\end{equation}
where $\beta(t)$ and $\rho(t)$ represent the inverse temperature and the monomer concentration at time $t$, and $E(N)$ is the potential energy of the fully grown targeted structure which is of size $N$. Note that we consider all component species corresponding to the targeted structure to be consumed at the same rate.



\begin{figure*}
\centering
\includegraphics[width=0.95\linewidth]{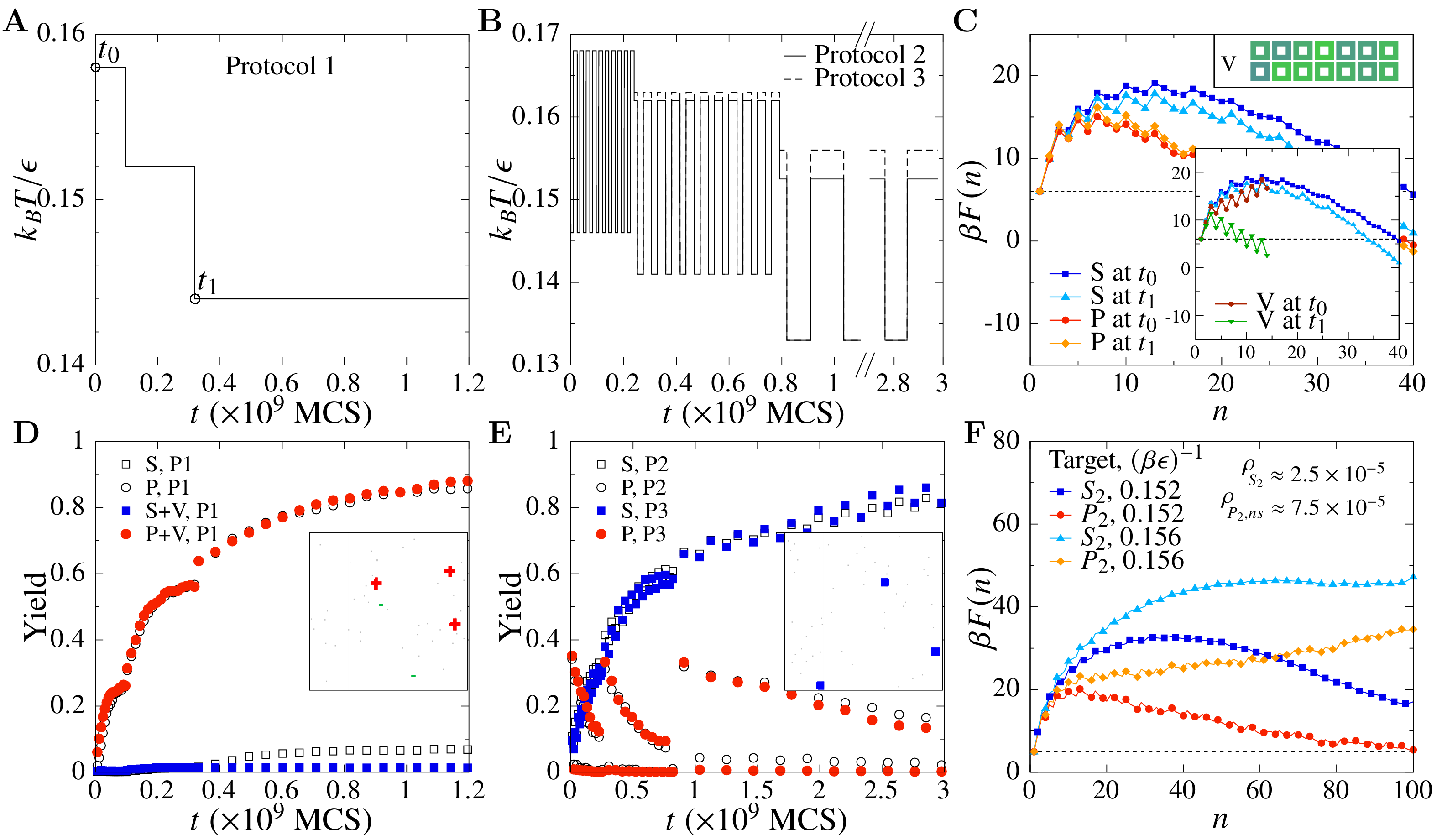}
\caption{\footnotesize Panels (A) and (B) show the protocols used to retrieve the targets P and S respectively. Simulations were done on $200 \times 200$ lattices with $3$ copies of each component. Both protocols have three sections, each lower in temperature than the previous, to compensate for monomer consumption (see Eq~(\ref{eq:rescaledtemp}) and text). As the targets only partly share components, growth of either differentially lowers the free energy of the other. Panel (C) shows this for protocol 1. The nucleation barrier of S is lower at time $t_1$ than at $t_0$. Hence, a fraction of S is also formed as shown by the open symbols in panel (D). Similarly for protocol 2, the free energy of P is lowered and it no longer melts completely as shown by the open symbols in panel (E). We create a vestigial aggregate V (top right corner, panel C) from the non-shared components of S (open green squares, Fig.~\ref{fig:optcomposition} A), with interactions such that it stabilizes faster than S (inset, panel C). V acts as a buffer improving selective retrieval of P as shown by the solid symbols in panel (D). For targeting S, a slight increase in the melting temperature (protocol 3, panel B) ensures that P is fully melted as shown by solid symbols in panel (E). The insets of panels D and E show typical configurations obtained at the end of protocols 1 and protocol 3, where aggregates matching the targets P, S and V are shaded in red, blue and green respectively. Panel (F) shows a case when two structures S$_2$ and P$_2$ share only one third of the component library ({\it SI Appendix}, Sec.~7). When two copies of S$_2$ are formed, the free monomer concentration of the components of S$_2$ is $\rho_{S_2} \approx 2.5\times 10^{-5}$, and that of the non-shared components of P$_2$ is $\rho_{P_2,ns} \approx 7.5 \times 10^{-5}$. This stabilizes P$_2$ at the expected melting temperature ($k_BT/\epsilon = 0.152$). The increased temperature required to melt P$_2$ ($k_BT/\epsilon = 0.156$) also destabilizes S$_2$.}
\label{fig:yieldedgeopt}
\end{figure*}

Fig.~\ref{fig:yieldedgeopt}(A) and (B) show the temperature protocols used to retrieve the targets P and S respectively. We perform the simulations on $200\times 200$ lattices with $3$ copies of each component type. Correspondingly, both the protocols have three sections, each lower in temperature than the previous section, so as to ensure that the free energy of the targeted structure remains invariant with monomer depletion, according to Eq.~(\ref{eq:rescaledtemp}). The depletion of monomers and hence the temperature shift is discrete in the present simulations given the small system size, but a more continuous variation according to Eq.~(\ref{eq:rescaledtemp}) will apply for larger system sizes, such as those relevant experimentally. 


\subsection*{Non-shared components are detrimental to selective retrieval}

Panels (D) and (E) of Fig.~\ref{fig:yieldedgeopt} show the yields of the two targets (open symbols) obtained with their corresponding protocols. The yield is computed as the fraction of particles that have assembled into copies of a given target structure, and the results are averaged over thermal fluctuations in a window of $10^5$ MCS and $96$ independent runs. The oscillations in the yield of P seen in panel (E) correspond to the formation and melting of P, during a temperature cycle. 
 
Although the yields of the targeted structures are high ($> 80\%$), there is small fraction of the other structure that is formed. As the targets only partly share the component library, targeting either leads to differential depletion of components. This lowers the free energy of the structure that is not being targeted. This is shown in Fig.~\ref{fig:yieldedgeopt}(C), for the case when P is targeted. At time $t_1$ indicated in panel (A), two thirds of the shared components are consumed, and the difference in the free energy barriers between S and P is lower than the initial value. This would not be the case if all components were depleted equally. As a result, some copies of S are nucleated as shown by the open squares in panel (D). This effect is more pronounced the fewer the fraction of components shared by the structures ({\it SI Appendix}, Sec.~7).

A similar effect occurs when targeting S, where nuclei of P are not completely melted as shown by the open circles in panel (E) of Fig.~\ref{fig:yieldedgeopt}. Thus a fraction of the nuclei of P are never recycled to form S. This can be rectified by slightly increasing the melting temperature, as shown by the dashed lines (protocol 3) in Fig.~\ref{fig:yieldedgeopt}(B). The corresponding yields are shown in solid symbols in panel (E). However, if the differential depletion leads to a significant lowering of the free energy of P, then the increase in temperature required to dissolve P might also dissolve S. This would happen if the two structures shared fewer components. We illustrate such a case in panel (F), and discuss further in {\it SI Appendix}, Sec.~7. These results demonstrate that good design requires the components to be maximally shared between the two structures.

\subsection*{Vestigial aggregation as a buffer for differential depletion}

In protocol 1, the nucleation barrier to S decreases as aggregation proceeds owing to the relative increase in the concentration of the non-shared components. 
However, we can mitigate this by programming attractive interactions between the non-shared components of S (the open green blocks in Fig.~\ref{fig:optcomposition}(A)), such that they form an additional or vestigial aggregate V faster than S can nucleate, thus acting as a buffer.

This is possible because the components from which we construct V are not consumed by P, whereas a majority of the species that constitute S are. Combined with the decreasing temperature, this allows us to choose the bond strengths such that at the start of protocol 1, V is unstable and with time it stabilizes with a much lower barrier than S. Such an aggregate is shown in the top right corner of Fig.~\ref{fig:yieldedgeopt}(C), and its free energy curves at times $t_0$ and $t_1$ are shown in the inset. It is constructed so as to not share bonds with any other structure composed of the vestigial components, to avoid chimeric aggregation, and with all bonds having strength $-1.11 \epsilon$. However, its edges have non-zero interactions as it is constructed from components internal to S. Also note that such a scheme is not effective for S, nor is it needed, as any such aggregate (as also P) will be melted during temperature cycling.

In Fig.~\ref{fig:yieldedgeopt}(D) we show the yields of the two structures with protocol 1 and in the presence of additional interactions encoding the vestige V (solid symbols). The aggregate V acts as a buffer improving selective retrieval of P. For the protocol S$'$, the presence of the additional interactions corresponding to V do not interfere with the nucleation of S. This is because we constructed S such that the components that form V (open green blocks, Fig.~\ref{fig:optcomposition}) are not at the core of S. With the updated protocol S$'$ and the vestigial structure, we find both good yields as well as selectivity. Finally, we note that the vestigial aggregation works well to buffer differential depletion even if a smaller fraction of components are shared ({\it SI Appendix}, Sec.~7), when targeting P.

\section*{Summary}

We have investigated the design of a multi-component mixture that is capable of forming more than one aggregate, and the ability to selectively target one of these structures through a suitable temperature protocol. 
Our results reveal that developing protocols for such addressable self-assembly crucially involves the design of the two structures, so as to prevent spurious or chimeric aggregates, and to maximize the selectivity of retrieval. 
We consider selective assembly of either one of two competing structures, and the generic situation wherein one of them has a lower free energy barrier to nucleation whereas the other structure has lower free energy upon complete assembly. 
We show that such a situation can be realised through an appropriate choice of interaction strengths of components. 
We show that the avoidance of spurious aggregation pathways favors a design where individual components have different bonding partners in the two structures. 
Our design also requires inclusion of some  components that are not shared by both  target structures, but we find that selectivity of assembly is maximally achieved when the components shared by both structures are maximised. 
Our results highlight the role played by secondary aggregation products, which we term {\it vestigial aggregates}. 
We demonstrate that with such design, protocols of temperature variation can be defined that selectively assemble the desired structures with high selectivity and yield. 
We believe these results provide valuable guidance for further experimental investigations of addressable self-assembly of multiple targets in multi-component self assembling systems.

\section*{Materials and Methods}
\subsection*{Lattice model for self-assembly}
We model our self-assembly mixture on a 2$d$ lattice of size $L \times L$, where each site can be empty or occupied by one of the $M$ possible components which are square blocks with four distinct interacting edges and orientations. Let $\rho_{i} = n_{i} / L^2$ be the concentration of the component ${i}$. We denote the interaction strength between two component edges $k,l \in \{1,2,3,...,4M\}$ by $I_{kl}$. Given a configuration of the system, its potential energy is given by
\begin{equation}
E = \sum_{\langle kl \rangle} I_{kl},
\label{eq:m2}
\end{equation}
where the sum is performed over all pairs of component edges that are in contact. We simulate the self-assembly dynamics using the virtual move Monte Carlo (VMMC)~\cite{Whitelam2009}, so as to enable the movement of clusters in simulations.

\subsection*{Data Availability}
All study data are included in the article and/or the SI Appendix.


\section*{Acknowledgements}

The authors acknowledge support from the UK-India Education and Research Initiative and the Department of Science and Technology, India, under Grant No.s IND/CONT/G/16-17/104, DST/INT/UK/P-149/2016, the Thematic Unit of Excellence on Computational Materials Science (TUE-CMS) and the National Supercomputing Mission facility (Param Yukti) at the Jawaharlal Nehru Centre for Advanced Scientific Research (JNCASR), for computational resources. SS acknowledges support through the JC Bose Fellowship (JBR/2020/000015) from the Science and Engineering Research Board, Department of Science and Technology, India.


\bibliography{multitarget-self-assembly.bib}

\end{document}



\title{Supplementary Information: Temperature Protocols to Guide Selective Self-Assembly of Competing Structures}

\author{Arunkumar Bupathy}
\email{bupathy@jncasr.ac.in}
\affiliation{Theoretical Sciences Unit, Jawaharlal Nehru Centre for Advanced Scientific Research, Bangalore, India.}

\author{Daan Frenkel}
\email{df246@cam.ac.uk}
\affiliation{Department of Chemistry, University of Cambridge, Cambridge, UK.}

\author{Srikanth Sastry}
\email{sastry@jncasr.ac.in}
\affiliation{Theoretical Sciences Unit, Jawaharlal Nehru Centre for Advanced Scientific Research, Bangalore, India.}

\maketitle


\section{Free Energy of Multi-Component Structures}
\label{sec:fe}

We compute the free energy of our multi-component target structures using the formalism of Jacobs \textit{et al.} \cite{Jacobs2015a, Jacobs2015}. To compute the free-energy difference between on-pathway aggregates of a particular size and the free monomers, we must consider all the ways in which a correctly bonded fragment can be assembled. Let $D(n,b)$ denote the set of all correctly bonded fragments of size $n$ monomers and $b$ bonds of a given structure, and $|D(n,b)|$ be the number of such fragments or the density of states. The free energy of an on-pathway aggregate of size $n$ monomers can be written as
\begin{equation}
\beta F(n) = -\ln \sum_{b} |D(n,b)| \bar{Z}_{n,b},
\label{eq:fe1}
\end{equation}
and the average fugacity $\bar{Z}_{n,b}$ of a fragment of size $n$ monomers and $b$ bonds is given by
\begin{equation}
\ln \bar{Z}_{n,b} = b \beta \tilde \epsilon_{n,b} + n \ln \rho - (n-1) \ln q_c.
\label{eq:fe2}
\end{equation}
Here, $\tilde \epsilon_{n,b}$ represents the mean bond energy of the fragment, $n \ln \rho$ represents the loss in entropy due to bringing $n$ monomers into contact, $\rho$ is the density of free monomers in the reservoir, and the last term represents the loss in rotational entropy with $q_c$ being the co-ordination number of the lattice.  The mean bond energy is given by
\begin{equation}
\beta\tilde\epsilon = \frac{1}{b} \ln \left\langle \exp \left(-\beta \sum_{i \in g} \epsilon_i \right) \right\rangle_{g \in D(n,b)}
\label{eq:fe3}
\end{equation}
where $\epsilon_i$ represent the energy of the bonds in a fragment $g$ in the set $D(n,b)$.

For each target structure, we make a connectivity graph $G$ where the vertices of the graph represent the individual components and the edges represent the bonds between them. Connected sub-graphs of $G$ represent correctly bonded fragments of the structure. We count the number of connected sub-graphs of size $n$ vertices and $b$ edges, statistically using the Wang-Landau flat-histogram algorithm~\cite{Wang2001}. We start with a randomly initialized sub-graph $g$ of $G$, and a flat $|D(n,b)| \equiv 1$ for all $n$ and $b$. At each step, we either add a randomly chosen vertex that is adjacent to the given sub-graph $g$ or remove a randomly chosen vertex from $g$. The removal of a vertex that breaks the sub-graph $g$ into disconnected parts (called cut-vertex or articulation point) is not allowed. The acceptance probability for transition from sub-graph $g$ to $g'$ is
\begin{equation}
p(g \to g') = min \left[1, \frac{|D(n(g),b(g))| n_{\pm}(g)}{ |D(n(g'),b(g'))| n_{\mp}(g')} \right]
\label{eq:fe4}
\end{equation}
where $n_{+}(g)$ is the number of vertices in $G$ that are adjacent to the sub-graph $g$, and $n_{-}$ is the number of non-cut-vertices in $g$.

Each time a state $(n,b)$ is visited, we update $|D(n,b)|$ by a multiplicative factor $f > 1$. We keep sampling the state space of connected sub-graphs, and updating $|D(n,b)|$ until we sample all states with roughly the same frequency. This procedure is repeated with smaller values of $f$, until we reach the desired level of accuracy. We then perform a biased Monte Carlo calculation using the density of states we just computed, with the acceptance probability of Eq.~(\ref{eq:fe4}), so as to compute the average bond energy $\bar{\epsilon}_{n,b}$.

\section{Effect of Maximizing the Shared Bonds}
\begin{figure*}
\centering
\includegraphics[width=0.95\linewidth]{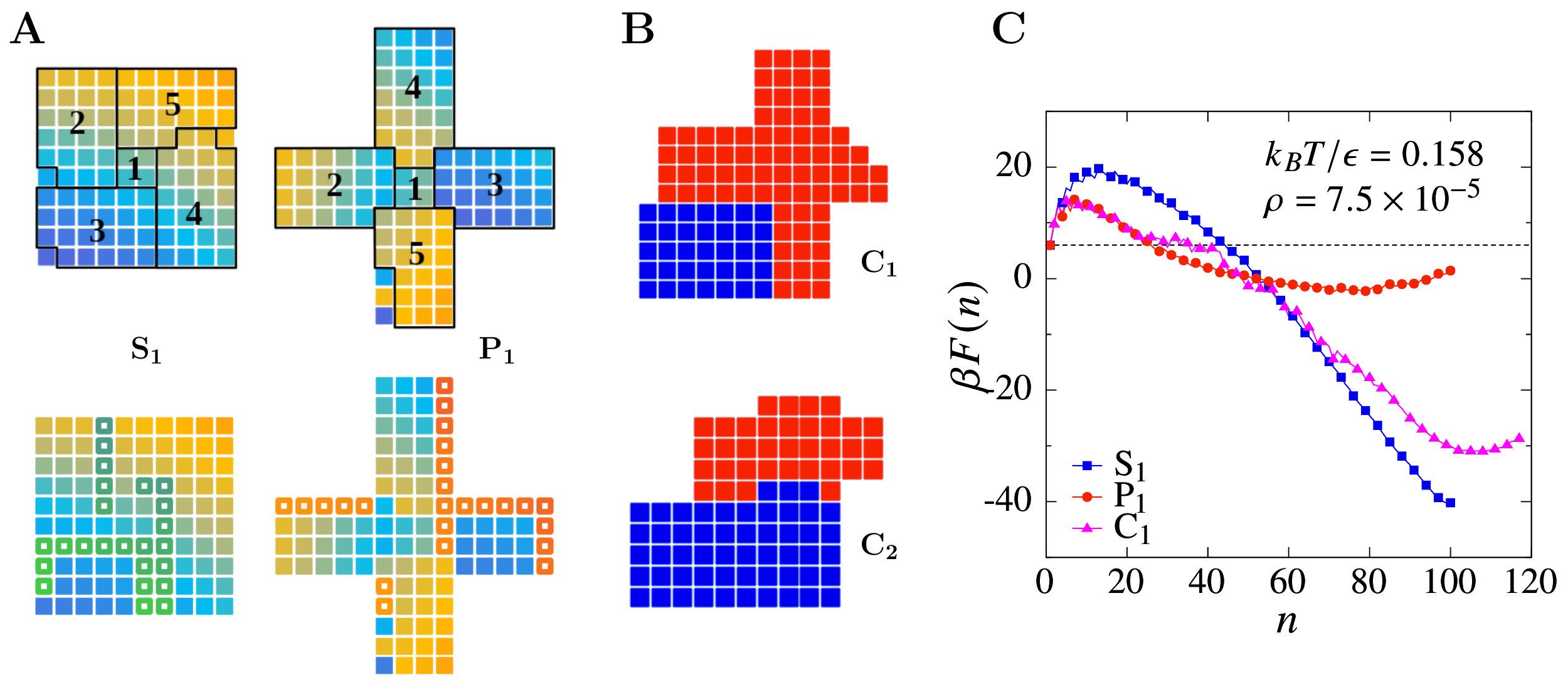}
\caption{Panel (A) shows a composition where the number of shared bonds between the two targets S$_1$ and P$_1$ have been maximized, leading to large chunks of one target being reused by the other as indicated by the numbered, outlined regions. There are some boundary species in either target that have non-zero interactions, as they are internal in the other. These are replaced with additional components to make the boundaries inert, as shown in the bottom of panel (A). Panel (B) shows the typical aggregates obtained with the protocol 1 shown in Fig.~3(A) of the main text. Panel (C) shows that the chimeric aggregate (C$_1$) has the same nucleation barrier as that of P. But since it is more stable than P, most aggregates turn into chimeras.}
\label{fig:shared}
\end{figure*}

We discuss here the effect of maximizing the number of shared bonds between the two targets structures that we consider. In other words, we try to construct the two targets such that most components have the same set of neighbors in both the targets. We start with an initial composition where both targets are made of the same $100$ distinct component species with random internal permutation of components. For the sake of clarity we call these S$_1$ and P$_1$ so as to distinguish from the ones in the main text. We then iterate on this so as to \textit{maximize} the cost
\begin{equation}
C = n_{shared-bonds}
\label{eq:cost}
\end{equation}
where $n_{shared-bonds}$ denotes the number of component edges that have the same neighbor (with the appropriate orientation) in the two targets. Note that each component has 4 edges and orientations, and we consider all $4M$ edges to be distinct. The left panel of Fig.~\ref{fig:shared} shows the composition of the two target shapes we consider in the main text, after minimizing $C$, leading to large chunks of one target being reused by the second as indicated by numbered outlined regions. As some edges of these pieces are internal to either targets, they must have non-zero interactions for the targets to be stable. This inevitably leads to parts of the boundaries of the two targets having non-zero interactions. To prevent aggregation at the boundaries, we replace the exposed boundaries with additional component species so as to passivate them. The resulting composition is shown in the bottom of Fig.~\ref{fig:shared}(A).

We then tune the bond strengths, as described in the main text, and perform simulations in the canonical ensemble, with the protocol 1 shown in Fig.~3(A) of the main text. In Fig.~\ref{fig:shared}(B) we show typical aggregates obtained at the end of the simulation run. In all cases, we obtain chimeric aggregates which are part S$_1$ and part P$_1$, as identified by the blue and red regions respectively. We identify one of the chimeric aggregates, and compute the free energy curve for the chimera. The free energies of S$_1$, P$_1$ and the chimera C$_1$ are shown in the panel (C) of Fig.~\ref{fig:shared}. The important point to note is that the chimera has a nucleation barrier that is comparable to P$_1$, which we are trying to grow. But the chimera is more stable than P$_1$. So in most cases we observe the chimera rather than the desired structure. Furthermore, we cannot melt the chimeric aggregate through temperature cycling, as that would destroy any nuclei of P$_1$.

\section{Spurious aggregation from bonded loops of incompatible components}

\begin{figure}[htb]
\centering
\includegraphics[width=0.95\linewidth]{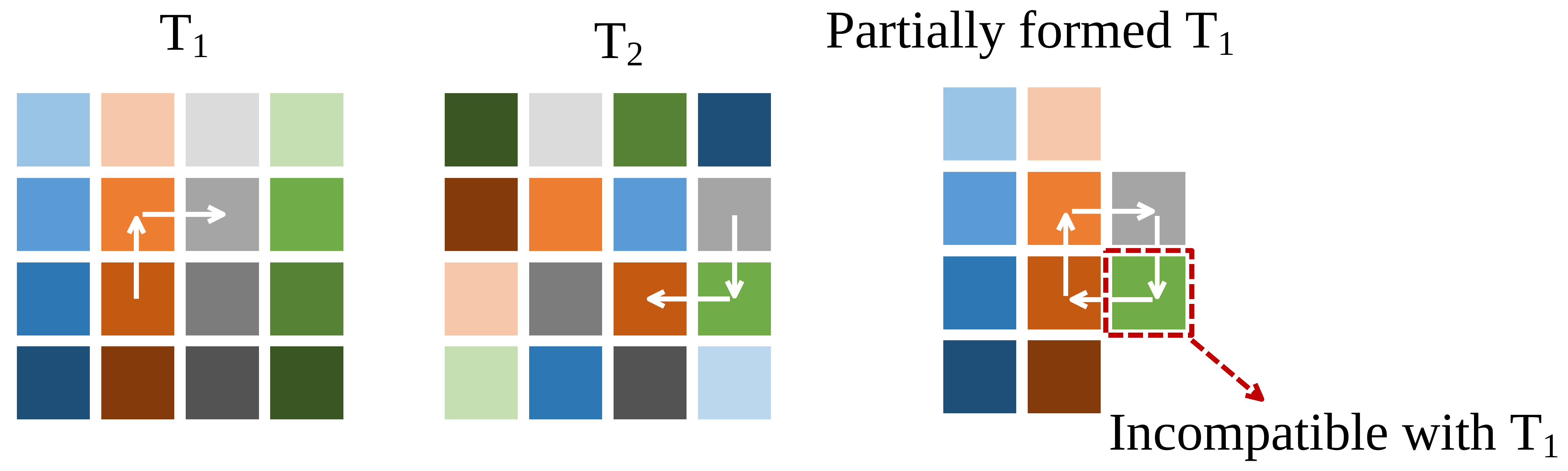}
\caption{Two example structures $T_1$ and $T_2$ that differ in their internal arrangement of components such that they have no common bonds. The rightmost panel shows an aggregate that is compatible with $T_1$ except for one monomer. The loop of bonds that stabilizes the incompatible monomer is not present in either $T_1$ or $T_2$, but the individual bonds are.}
\label{fig:incompatible_loops}
\end{figure}

In the main text, we show how chimeric aggregation can occur if the two targets share the same bond, i.e., a pair of neighboring components. However, even with no shared bonds, there is a possibility of incompatible components forming bonded loops. Such erroneous loops are harder to correct as they require the breaking of two bonds. We illustrate this in Fig.~\ref{fig:incompatible_loops}. On the left we show two example structures $T_1$ and $T_2$ with different internal arrangements such that there are no common bonds between them. On the right, we show a partially complete aggregate that is compatible with $T_1$, except for one block as highlighted by the broken red rectangle. This incompatible piece is stable against removal as it is held together by two bonds. The complete loop of bonds that stabilizes the incompatible monomer is not present in either $T_1$ or $T_2$, but the individual bonds are.
In general, any such loop irrespective of its size can present an opportunity for erroneous aggregation. But the most important loops are those made of four nearest neighbors, as longer loops are much less likely to form. If present in large numbers such loops could also cause chimeric aggregation. However, this is only an issue when the number of stored structures is large or the component library is small. We therefore do not optimize the target compositions against such incompatible loops.

\section{Tuning the free energy landscapes manually}

\begin{figure*}
    \centering
    \includegraphics[width=0.9\linewidth]{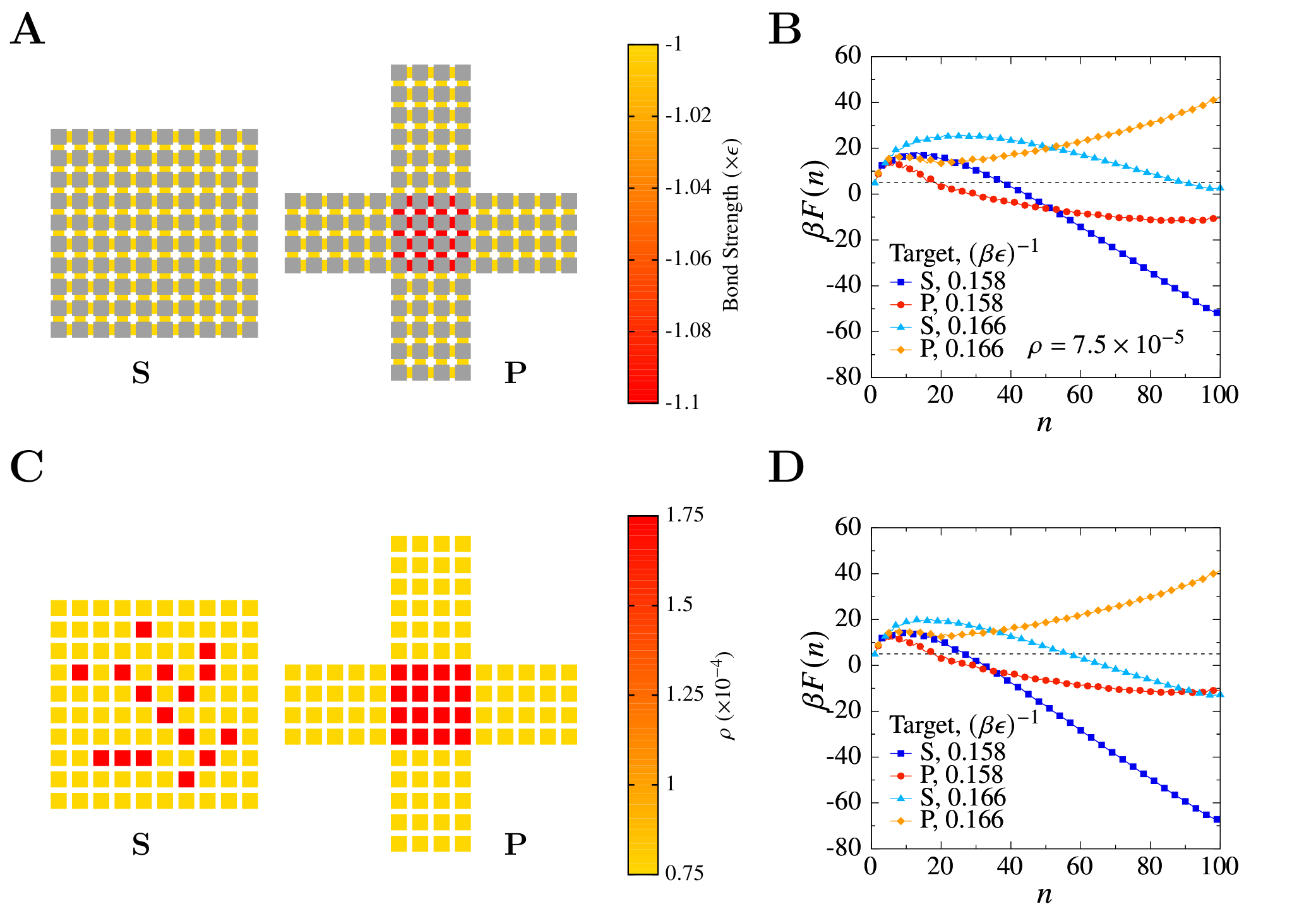}
    \caption{Panel A shows the strength of the interactions in the two structures S and P. A $4\times 4$ region in the core of P is given stronger bonds. This reduces the nucleation of barrier of P relative to S as shown in panel B. Panel C shows the free monomer concentration of the component species that constitute the two structures, where the components at the core of P are enriched. Panel D shows the free energy curves as a result of the concentration pattern.}
    \label{fig:manualtune}
\end{figure*}

In the main text, we considered the generic case in which one of the target structure has a lower free energy barrier to nucleation whereas the other structure has a lower free energy minimum upon complete aggregation. The manner in which the design of aggregates in such a case could we optimised was described. Here, we demonstrate that target structures can indeed be prepared to satisfy such an assumption. To do so, we start by considering the initial structures considered in the main text, with free energy curves as shown in Fig.~1(B), displaying the same free energy barrier for both structures at the lower temperature shown, and with the S structure having a lower free energy upon complete formation.
The free energy barrier of the P structure can be lowered by selecting a region at its centre, which is about the size of the critical nucleus ($\sim 4\times 4$) and increasing the strength of the bonds within the region, or by increasing the concentration of the species in that region.
We need not optimize for the free energy minima since the free energy minimum of P is already higher than S. This is illustrated in Fig.~\ref{fig:manualtune} where the panels A and B show the effect of varying the bond strengths and panels C and D show the effect of varying the chemical potentials. Note that the free energies computed using the procedure outlined in Sec. \ref{sec:fe} assumes that intermediate sized clusters of a given size $n$ are equivalent regardless of composition. This assumption is not strictly valid in the present case, and therefore a more careful computation of the free energy is necessary if the precise values of the barrier height are of interest. We do not pursue such a calculation here, since our purpose is simply to demonstrate that the free energy profiles can be modified in desired ways by tuning the interaction energies or chemical potentials. 

Since the bonds that constitute the $4\times 4$ region in P are not shared by the two structures, varying the bond strength of the corresponding bonds does not affect the bonds in S. On the other hand, since the same components are shared between the two structures, increasing the concentration of the species constituting the $4\times 4$ region in P will also lower the free energy curves of S, but to a smaller degree near the free energy barrier, as these components are not spatially correlated in S.

\section{Computing the Nucleation Rates}

We need to know the nucleation rates of the target structures to identify the length of the simulation protocols. We follow Debenedetti~\cite{Debenedetti_1996} to derive the nucleation rates for the multicomponent self-assembly system. Consider the system in a supercooled state, where the gaseous phase is metastable. As a result of density fluctuations, small nuclei are constantly being formed and destroyed. We assume that the concentration of these nuclei is small, and that they grow or shrink by one particle at a time. We can then write the difference between the rate at which $n$-particle nuclei are formed and the rate at which they are destroyed, due to single particle events as
\begin{equation}
J(n) = f(n-1)P(n-1)\gamma(n-1) - f(n)P(n)\alpha(n)
\label{eq:nuc1}
\end{equation}
where $f(n)$ is the concentration of $n$-particle nuclei; $P(n)$ represent their perimeter; $\gamma(n-1)$ is the flux per unit time and length (area in 3$d$) of single particles onto an $(n-1)$-particle nucleus; and $\alpha(n)$ is the flux of single particles leaving an $n$-particle nucleus. Of the two rates $\gamma$ can be computed, but $\alpha$ is not known. Assuming that an equilibrium distribution of nuclei can be established in the bulk metastable phase, we can write
\begin{equation}
C(n-1)P(n-1)\gamma = C(n)P(n)\alpha
\label{eq:nuc2}
\end{equation}
where we have assumed that the fluxes are independent of the size of the nuclei, and $C(n)$ is the equilibrium concentration of nuclei of size $n$. Using (\ref{eq:nuc2}) in (\ref{eq:nuc1}), we get
\begin{equation}
J(n) = \gamma P(n-1)C(n-1)\left[\frac{f(n-1)}{C(n-1)} - \frac{f(n)}{C(n)} \right]
\label{eq:nuc3}
\end{equation}

For a time invariant propulation of clusters we should have
\begin{equation}
\frac{\partial f(n,t)}{\partial t} = J(n) - J(n+1) = 0,
\label{eq:nuc4}
\end{equation}
and is established when $J$ becomes independent of $n$. Rearranging (\ref{eq:nuc3}), and summing from $n=2$ to $n=\Lambda$, where $\Lambda$ is much larger than the critical nucleus size, we obtain
\begin{equation}
J = \frac{\frac{f(1)}{C(1)} - \frac{f(\Lambda+1)}{C(\Lambda+1)}}{\sum_{n=1}^{\Lambda}\frac{1}{\gamma P(n)C(n)}} \approx \left[\sum_{n=1}^{\Lambda}\frac{1}{\gamma P(n)C(n)}\right]^{-1},
\label{eq:nuc5}
\end{equation}
where we have made use of the fact that in the bulk metastable phase the equilibrium and actual concentration of single particles are nearly the same and that at the onset of the phase transformation the concentration of sufficiently large nuclei is vanishingly small.

The equilibrium concentration $C(n) = \rho^{tot} \exp[-\Delta F(n) / k_B T]$, where $\rho^{tot}$ is the total monomer concentration, and $\Delta F(n) = F(n) - F(1)$ which we already compute. Here we have assumed that all $M$ components are present in equal concentration $\rho$ in the mixture, such that $\rho^{tot} = M \rho$. However, in our multi-component system not all incoming particles can form bonds with the nucleus. This is because each block in a given target structure is a distinct species with specific and directional bonds. Therefore, of the total flux ($\gamma^{tot}$) of incoming particles, only a fraction ($P(n)/M$) can form bonds with an $n$-particle nucleus, and with a probability $1/4P(n)$ since the bonds are specific and directional. So the effective value of $\gamma$ is
\begin{equation}
\gamma = \frac{1}{4 M}\gamma^{tot}
\label{eq:nuc6}
\end{equation}

If the monomers are diffusing with a diffusion constant $D$ and considering a square nucleus of size $n=4r^2$, we can approximate $\gamma^{tot} = \rho^{tot}4r^2/t \approx \rho^{tot}8Dd$ and $P(n) \approx 4\sqrt{n}$ which gives
\begin{equation}
J \approx \frac{8Dd(\rho^{tot})^2}{M} \left[\sum_{n=1}^{\Lambda} \frac{\exp[\Delta F(n)/k_B T]}{\sqrt{n}} \right]^{-1}
\label{eq:nuc7}
\end{equation}

\begin{figure}
\centering
\includegraphics[width=0.75\linewidth]{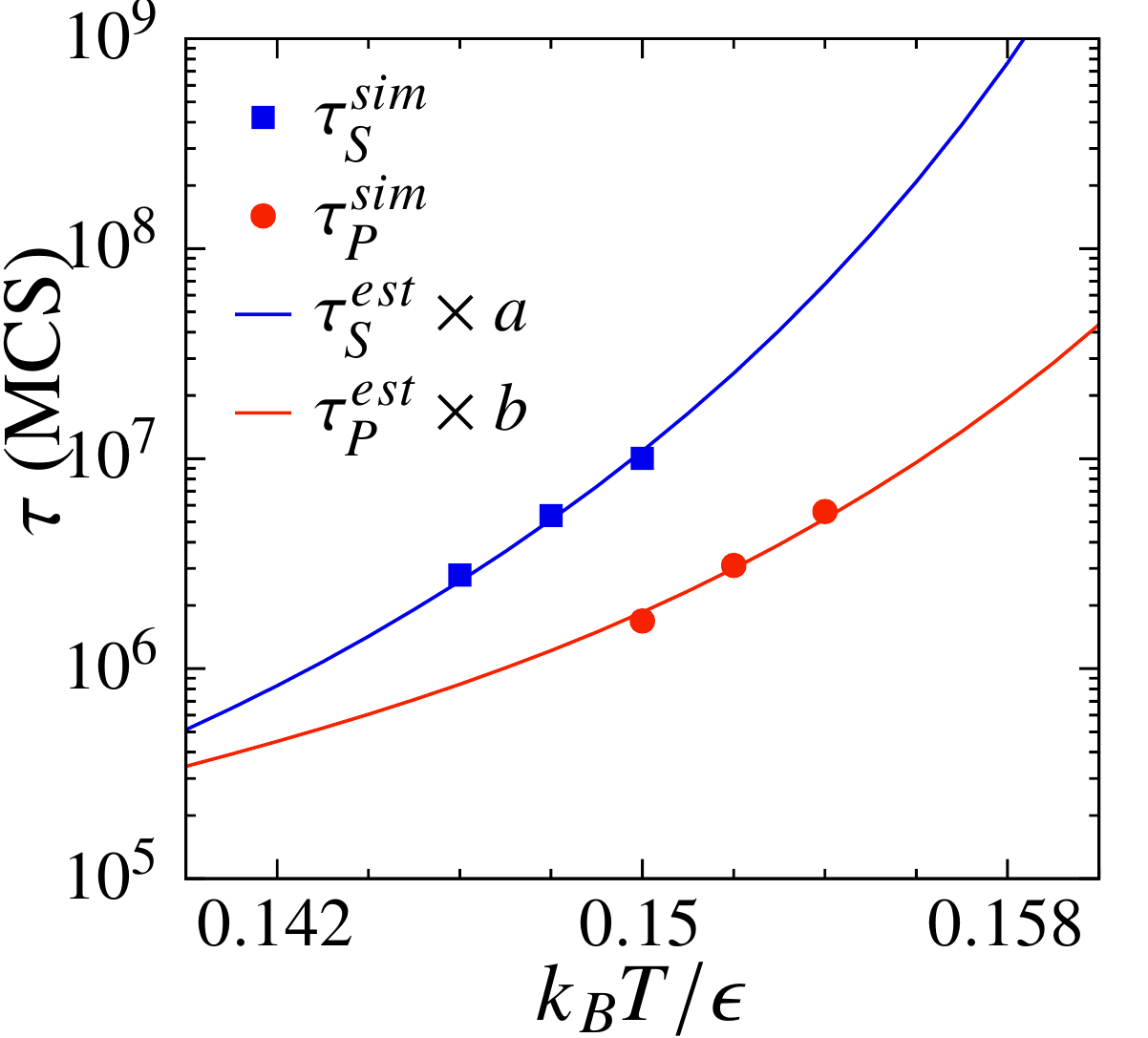}
\caption{Nucleation times from MC simulations (symbols) vs. estimated values (lines) for the two target structures. The estimated values for S and P are scaled by constant factors $a \approx 50$ and $b \approx 90$ to match the observed values.}
\label{fig:taus}
\end{figure}
In Fig.~\ref{fig:taus} we show the nucleation times measured from MC simulations as well as the estimates from Eq.~(\ref{eq:nuc7}) scaled by constant factors. The nucleation times in simulations are measured as $2 \times \tau_{fp}(n^*)$, where $\tau_{fp}(n^*)$ is the time at the inflexion point of the mean first-passage time data. While the temperature dependence from MC simulations and the calculation above agree, the calculated rates are different by scale factors which are different for S and P, being $ \approx 50$ and $ \approx 90$ respectively. We use the scaled analytical values to extrapolate the nucleation rates to arbitrary temperatures and free monomer concentrations. 

\section{Scaling the temperature with monomer concentration}

Since we perform simulations in the canonical ensemble, growth of structures depletes free monomers. The rate at which the monomers are consumed can be written as
\begin{equation}
    \frac{\mathrm{d}\rho}{\mathrm{d}t} = -k \rho,
\end{equation}
where $\rho$ is the free monomer concentration of the components in the mixture, and the kinetic factor
\begin{equation}
    k = A e^{-\beta F(n^*)},
\end{equation}
where $F(n^*)$ is the nucleation barrier of the targeted structure, with a critical nucleus of size $n^*$ at an inverse temperature $\beta$. We want $k$ to remain constant, even while $\rho$ changes so that
\begin{equation}
    \rho = \rho_0 e^{-kt}
\end{equation}
where $\rho_0 = \rho(0)$ is the initial monomer concentration. For this to be true, we require
\begin{equation}
    \beta(t_1)F(n^*,\rho(t_1)) = \beta(t_2)F(n^*,\rho(t_2)) = const.
    \label{eq:constbarrier}
\end{equation}
If we assume 
\begin{equation}
    \frac{\beta(t_1)F(n^*,\rho(t_1))}{\beta(t_2)F(n^*,\rho(t_2))} = \frac{\beta(t_1)F(N,\rho(t_1))}{\beta(t_2)F(N,\rho(t_2))}
\end{equation}
and from Eq.~(\ref{eq:fe1}) and (\ref{eq:fe2}) we can write
\begin{align}
\beta(t)F(N,\rho(t)) &= \beta(t)E(N) + N\ln\rho(t) - (N-1)\ln q_c \nonumber \\
                     &= \beta(t)E(N) + N [-kt + ln\rho_0] - (N-1)\ln q_c \nonumber \\
                     &= \beta(0)F(N,\rho_0) \nonumber \\
                     &= \beta(0)E(N) + N\ln\rho_0 - (N-1)\ln q_c
\end{align}
This implies that $\beta(t)E(N) - Nkt = \beta(0) E(N)$ which gives
\begin{equation}
    \beta(t) = \beta(0) + \frac{N}{E(N)} kt = \beta(0) + \frac{N}{E(N)} \ln\frac{\rho_0}{\rho(t)}
\end{equation}


\section{Effect of sharing a smaller fraction of the component library}



\begin{figure*}
    \centering
    \includegraphics[width=0.98\linewidth]{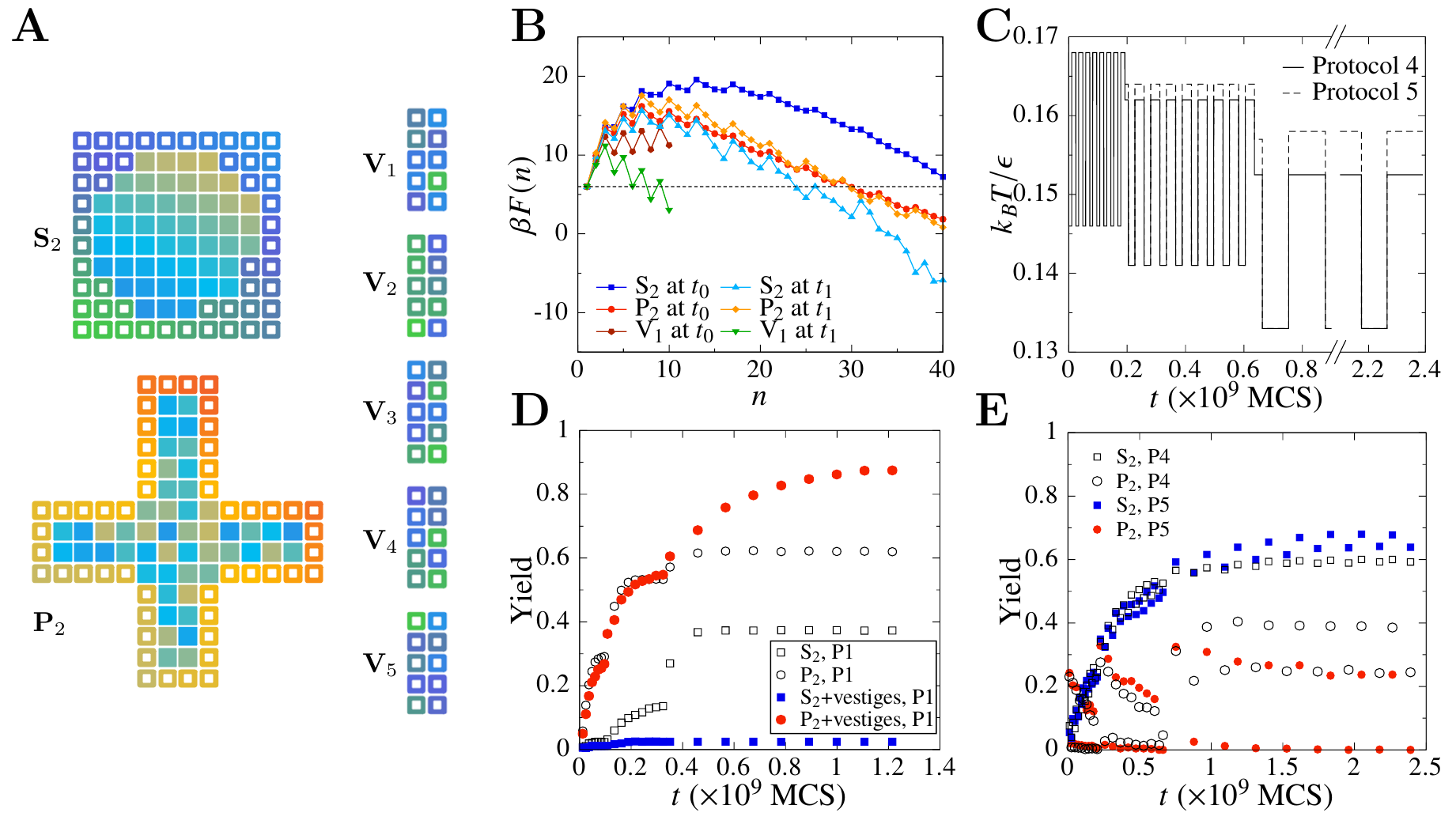}
    \caption{Panel (A) shows two structures S$_2$ and P$_2$ that share only 50 components (solid squares). The open squares of green-blue and orange-red shades are unique to either. The vestiges are constructed from the non-shared components (open blocks) of S$_2$. In the absence of vestiges, differential consumption of components by either target lowers the free energy of the other, as shown in panel (B) at two different times along protocol 1 of Fig.~3(A) in the main text. This lowers the yield of P$_2$ and a significant fraction of S$_2$ is also nucleated as shown by the open symbols in panel (D). With the vestiges, the yield of P$_2$ is significantly increased as shown by the solid symbols. Using protocol 4 shown in panel (C), which is constructed expecting all components to be consumed equally, a significant fraction of P$_2$ is never melted during the temperature cycling (open circles, panel (E)). Increasing the temperature (protocol 5, panel (C)) to melt fragments of P$_2$ also melts the nuclei of S$_2$, thus reducing the yield of S$_2$ as shown by solid symbols in panel (E).}
    \label{fig:shareless}
\end{figure*}

In Fig.~\ref{fig:shareless} (A), we show two structures S$_2$ and P$_2$, which are of the same shape and size ($N=100$) as those in the main text, but they share a smaller fraction of the component library. Specifically, they only share 50 components (solid squares) which are internal. The components that compose each of their boundaries (open squares) are unique to either structure. Thus the total component library is of size 150. The permutation of components is optimized as described in the main text, and we tune the free energy curves of S$_2$ and P$_2$, as described in the main text. For retrieving P$_2$, we use the same protocol 1 as shown in Fig.~3(A) of the main text, and protocol 4 shown in panel (C) for S$_2$. The differential consumption of monomers by the either target lowers the free energy of the other structure. The free energies of the different structures, at times $t_0$ and $t_1$ of protocol 1 (Fig.~3 (A), main text) are shown in panel (B). The free energy barrier of S$_2$ is lower than that of P$_2$, the targeted structure, at time $t_1$.

A similar effect is observed for S$_2$ as well (Fig.~3(F), main text). As the simulation proceeds, and two copies of S$_2$ are formed, the free energy curve of P$_2$ is shifted below S$_2$ at the temperature ($k_BT/\epsilon = 0.152$) where we expect P$_2$ to be unstable. This leads to the formation of the undesired structure, and lowers the yield of the desired one as shown by the open symbols in panel (D) and (E) of Fig.~\ref{fig:shareless}. The increase in temperature required to melt P$_2$ also destabilizes S$_2$, and hence the yield of $S_2$ is limited. as shown by the solid symbols in Fig~\ref{fig:shareless}(E). Also note that the melting temperature for P$_2$ in simulations can be slightly higher ($k_BT/\epsilon = 0.158$) than that predicted from the free energy curves ($k_B T / \epsilon = 0.156$, as shown in Fig.3(F) of the main text).

With additional vestigial aggregates shown in panel (A) constructed from the non-shared components of S, the selectivity of P$_2$ can be improved significantly. These vestiges are constructed so that they do not share any bonds with S$_2$. But note that their boundaries can have non-zero interactions, as some of them are internal in S$_2$. These aggregates nevertheless buffer the differential depletion of components and improve the yield of P$_2$ as shown by solid symbols in Fig.~\ref{fig:shareless}(D).



\bibliography{supplement}